\documentclass[11pt,letter]{article}
\usepackage[margin=1in]{geometry}

\usepackage{lipsum} 
\usepackage{graphicx}
\usepackage{amsmath}
\usepackage{caption}
\usepackage{subcaption}
\usepackage{cuted}
\usepackage{hyphenat}
\usepackage{setspace}
\usepackage{color,soul}

\begin{document}

\title{\textbf{Canister valve and actuator deposition in metered dose inhalers formulated with low-GWP propellants} }
\author{Daniel J Duke$^1$, 
Lingzhe Rao$^1$, 
Alan Kastengren$^2$,
Benjamin Myatt$^3$,\\
Phil Cocks$^3$,
Stephen Stein$^4$,
Nirmal Marasini$^5$,
Hui Xin Ong$^{5,6}$,
Paul Young$^{5,6}$
\vspace{5mm}\\
\small $^1$ Laboratory for Turbulence Research in Aerospace \& Combustion (LTRAC),\\
\small Department of Mechanical \& Aerospace Engineering, Monash University, Australia\\
\small $^2$ X-ray Science Divison, Argonne National Laboratory, Lemont, Illinois USA\\
\small $^3$ Kindeva Drug Delivery, Loughborough, United Kingdom\\
\small $^4$ Kindeva Drug Delivery, Woodbury, Minnesota USA\\
\small $^5$ Woolcock Institute of Medical Research, Glebe NSW 2037 Australia\\
\small $^6$ Department of Marketing, Macquarie Business School,\\ \small Macquarie University, Sydney, NSW 2109, Australia
}
\date{Preprint - \today}
\maketitle
\section*{Abstract}
\doublespacing
A challenge in pressurised metered-dose inhaler formulation design is management of adhesion of the drug to the canister wall, valve and actuator internal components and surfaces, especially for sedimenting or creaming suspensions. Visual analysis of drug solubility and suspension behavior is typically performed in transparent vials. If the results are affected strongly by wall-material interactions, they may not be replicable in typical metal canister pMDI systems. This is of particular concern in low-greenhouse warming potential (GWP) propellant formulations where the chemistry of the new propellants and solubility with many drugs is not yet fully understood. The same can be extended to deposition in  actuator and valve components, which are generally opaque. In this study, we demonstrate a novel application of X-ray fluorescence spectroscopy using synchrotron radiation to assay the contents of surrogate solution and suspension pMDI formulations of potassium iodide and barium sulfate in propellants HFA134a, HFA152a and HFO1234ze(E) using standard components. Through unit life drug distribution in the canister valve closure region and actuator can vary more significantly with new propellants. These effects must be taken into consideration in the development of products utilising low-GWP propellants.

\section{Introduction}

One of the major challenges in formulating pressurised metered dose inhaler (pMDI) products is managing the sedimentation and creaming behaviour of suspended drug in the canister, and how quickly these processes occur after agitation \cite{1992.Byron,Hatley.2017}. In addition, `wall losses' can occur in both suspension and solution products due to deposition of the active pharmaceutical ingredient (API) or drug onto the canister and valve components over time due to physical or chemical interactions \cite{Young.2003}.  The strong electronegativity of the hydrofluoroalkane (HFA) propellants used in many pMDIs also exacerbates this effect \cite{1999.Vervaet}. The problem has been quite successfully managed for current-generation HFA pMDIs through the use of surfactants \cite{Stein.2014} and low surface energy coatings such as FEP (fluorinated ethylene propylene) and other fluoropolymers \cite{Ashayer.2004,Brouet.2006,Skemperi}. 

\begin{figure}[b]
\centering
\begin{subfigure}[b]{.32\textwidth} \centering
	\includegraphics[height=2.1cm]{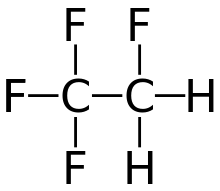}\caption{ HFA134a\\ \textit{GWP=1300}}
\end{subfigure}
\begin{subfigure}[b]{.32\textwidth} \centering
	\includegraphics[height=1.8cm]{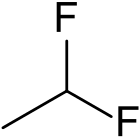}\caption{ HFA152a\\ \textit{GWP=138}}
\end{subfigure}
\begin{subfigure}[b]{.32\textwidth} \centering
	\includegraphics[height=1.6cm]{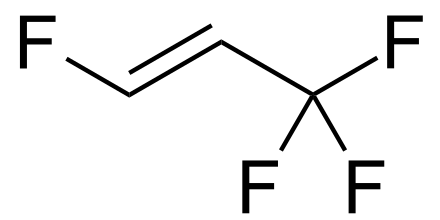}\caption{ HFO1234ze(E)\\ \textit{GWP$<$1}}
\end{subfigure}
\caption{Propellants considered in this study. GWP data from Pritchard \cite{Pritchard.2020}\label{fig1}.}
\end{figure}

The introduction of new low greenhouse warming potential (low-GWP) propellants such as HFA152a and HFO1234ze(E) (Figure \ref{fig1}) pose new challenges to pMDI formulation design \cite{Stein.2021}.  The low density of HFA152a \cite{1987.Sato} will significantly impact the suspension behavior of particles in a propellant based formulation, for example  \cite{Lewis.2023}.  HFO1234ze(E) has a similar density to HFA134a \cite{Katsuyuki.2016} which will aid in the reformulation of suspensions. However, the double bond which affords it a low GWP alters its electronegativity and may lead to drug losses on the canister wall and deposition on valve components.

Interactions between suspended or dissolved drugs and new propellants are not yet fully understood. The new propellants have significant differences in dipole moment and polarizability compared to current HFA propellants (see Table \ref{propellantPropertyTable})  \cite{Sampson.2019,Meyer.1991}. Such changes are likely to be better indicators of their drug interactions than vapor pressure or density, which are enhanced by the F-F repulsive forces which give HFAs and HFOs their favourable properties as propellants \cite{1999.Vervaet,Smyth.2003}. 

Regions of the internal canister volume with greater geometric complexity are preferable sites for drug deposition owing to both increased available surface area and the difficulty in coating these regions. This also applies to the canister, canister-valve closure, valve surfaces and retaining cup (if present) \cite{Stein.2014}. Coatings on valve components can be effective, but are not as commonly used as can coatings \cite{Jinks.2014}. It is possible that deposition on these components may also be exacerbated in low-GWP formulations.

\begin{table}
\centering
\resizebox{\textwidth}{!}{%
\begin{tabular}{lllllll}
\textbf{Propellant}&\textbf{Molar} & \textbf{Acentric} & \textbf{Dipole} & \textbf{Dielectric} & \textbf{Vapor} & \textbf{Liquid} \\
& \textbf{weight} & \textbf{factor} & \textbf{moment} & \textbf{constant} & \textbf{pressure} & \textbf{density} \\
&&&&\multicolumn{3}{c}{\textbf{\textit{(Saturated liquid at 25$^\circ$C)}}}\\ 
\hline
\textbf{HFA134a}& 102.032 & 0.327 & 2.06 $D$ & 10 & 6.7  & 1207 \\
\textit{1,1,1,2-tetrafluoroethane} &g mol$^{-1}$&&&&atm&kg m$^{-3}$\\
Refs. \cite{Meyer.1991,2007.Lemmon,Gbur, coolprop} \\
\hline
\textbf{HFA152a} & 66.05 & 0.275 & 2.26 $D$ & 14 & 5.6  & 904 \\
\textit{1,1-difluoroethane} &g mol$^{-1}$&&&&atm&kg m$^{-3}$\\
Refs. \cite{Meyer.1991,1987.Sato,2007.Lemmon,ppa, coolprop}\\
\hline
\textbf{HFO1234ze(E)}  & 114.043 & 0.313 & 1.13 $D$ & 7.5 & 4.6  & 1170  \\
\textit{1,3,3,3-tetrafluoropropene}&g mol$^{-1}$&&&&atm&kg m$^{-3}$\\
Refs. \cite{Sampson.2019,Katsuyuki.2016, 2007.Lemmon, coolprop}\\
\hline
\end{tabular}%
}
\caption{Selected chemicophysical properties of the propellants considered in this study.\label{propellantPropertyTable}}
\end{table}


\textit{In situ} assessment of drug deposition and sedimentation in metal canisters is limited due to a lack of optical or physical access to the internal surfaces during use. Fundamental particle-wall interactions can be assessed \textit{in vitro} using atomic force microscopy \cite{Young.2003} and colloidal probe microscopy \cite{Traini.2006}, but such highly sensitive measurements are not practical for HFA formulations \textit{in situ}. Most measurements involve replacing the metal canister with a transparent glass or plastic vial and observing bulk sedimentation or creaming behaviour \cite{DSa.2015}. These include, for example, laser reflectance \cite{Michael.2001}, infrared transmission (OSCAR) \cite{Brindley.1999}, combined optical transmission and backscatter (i.e. Turbiscan) \cite{Blanes.2022}, shadowgraphy \cite{Wang.2018nub,Moraga-Espinoza.2019} and manual observation \cite{Dellamary.2000,Abdrabo.2019}. Effects which depend upon the canister material or closure geometry will not be observed using these methods. Drug deposition on opaque valve components may not be easily observed, leaving only pMDI dose content uniformity and intrusive destructive inspection as indicators of these effects \cite{DSa.2015}.

An alternative approach to non-invasive, non-destructive \textit{in situ} measurement in metal canisters is the application of X-rays. X-ray computed tomography has been applied to individual particles at the microscale \cite{DSa.2015} and to canister and valve components at the macroscale \cite{White.2022}. There are only a limited number of studies investigating drug deposition using X-rays \cite{CocksSlowey}. Although techniques such as X-ray powder diffraction are frequently used to study spray-dried particles \cite{Tarara.2004}, the presence of the canister wall limits the range of useful X-ray photon energies to higher values which do not interact strongly with organic compounds in most APIs.

A solution to the problem of requiring higher X-ray photon energies to penetrate the canister wall is the use of X-ray fluorescence spectroscopy (XFS) with a high atomic number marker. In prior work, we have shown that the bromine atom in Ipratropium bromide can be used as a naturally-occurring marker for measurements in pMDI sprays using a focused synchrotron beam at 15 keV \cite{2015.Duke,2019.Duke5w7}. XFS is tolerant of much lower concentrations than absorption-based measurements \cite{2011.Kastengren}. However, measurements in aluminium canisters require higher energies than those in sprays ($E_0 \geq$ 40 keV) and thus higher atomic number markers (i.e. I, Xe, Cs, Ba) \cite{2009.Thompson}. For example, prior X-ray fluorescence measurements in aluminium nozzles have utilised micronised cerium particles \cite{2018.Duke}.

In this study, we consider two surrogate molecules. A solution of potassium iodide (KI) in ethanol blended with HFA or HFO propellant (Figure \ref{fig1}) is used as a model solution formulation, with the iodine atom acting as the X-ray fluorescent marker. A suspension of micronised barium sulfate (BaSO$_4$) in pure HFA or HFO propellant is used as a model suspension formulation. We use these models to demonstrate the efficacy of the X-ray fluorescence technique for canister, valve and actuator deposition measurements in standard pMDI hardware. The primary advantage of this technique is the ability to sample inside the canister, valve and actuator regions non-destructively with realistic API concentrations, and to therefore be able to assess through-life performance by scanning the pMDI between repeated firing events.

\section{Methods}

\subsection{Formulations}

In this study, a total of six formulations were considered; one model solution and one model suspension at fixed concentration of nominally 2 mg/mL in three propellants. For the model solutions, potassium iodide (Sigma-Aldrich) was dissolved in absolute ethanol (Sigma-Aldrich) to form a concentrate using a magnetic stirrer. Weighing of all APIs was conducted on a Laboratory balance (AS 62.R2+, Radwag) with $\pm 0.02$ mg precision.
The concentrate was dispensed into the canisters to predetermined weight prior to crimping with a valve and pressure-filling with
 one of either HFA134a, HFA152a or HFO1234ze(E) propellant (industrial grade, A-Gas Australia, 99.5\% purity)
using a Pamasol (Laboratory plant 2002, Pamasol, Switzerland) to 200 nominal actuations.

\begin{figure}
\centering
\begin{subfigure}[b]{.54\textwidth} \centering
	\includegraphics[width=\textwidth]{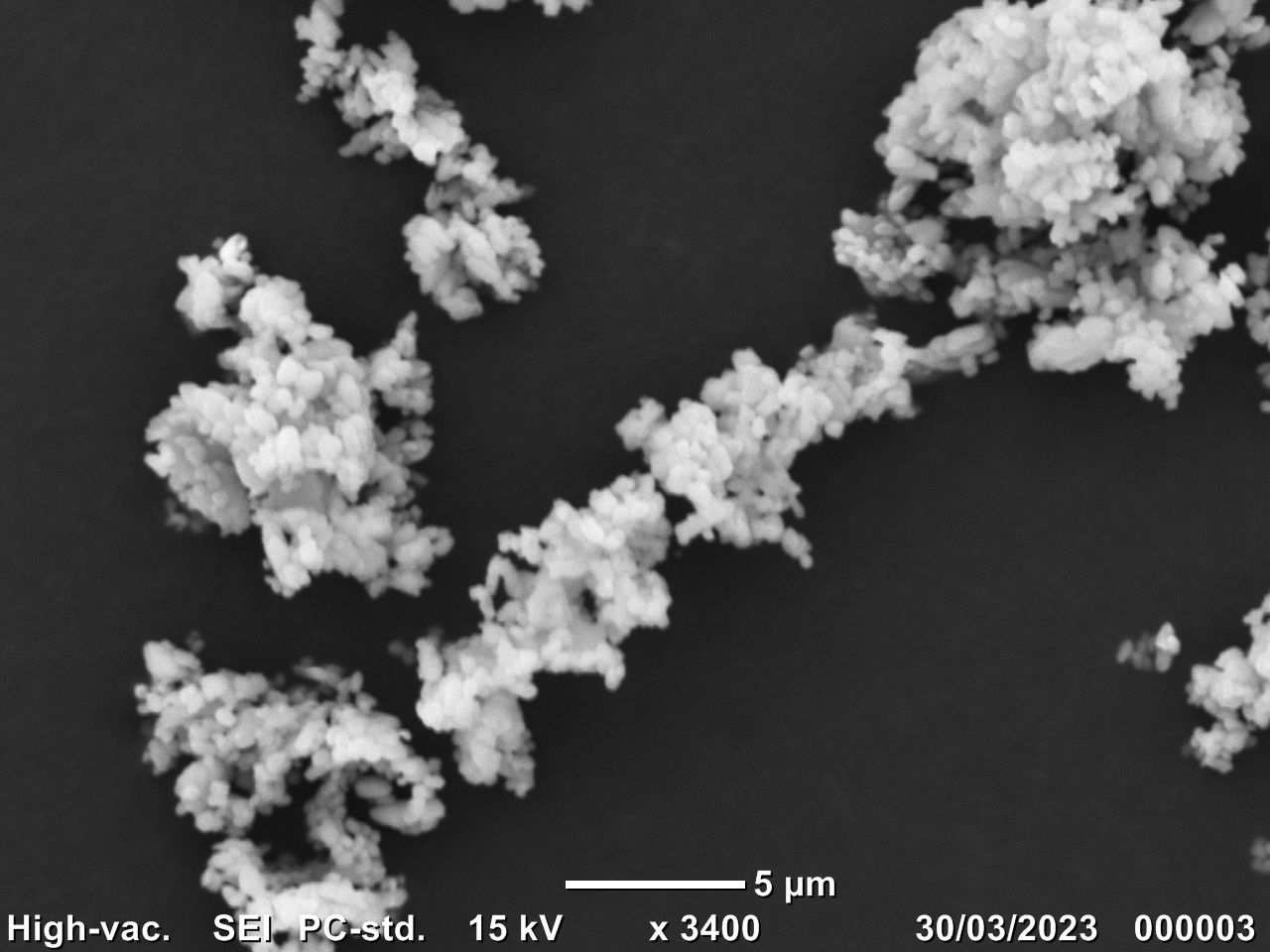}\caption{SEM image.}
\end{subfigure}
\begin{subfigure}[b]{.45\textwidth} \centering
	\includegraphics[width=\textwidth]{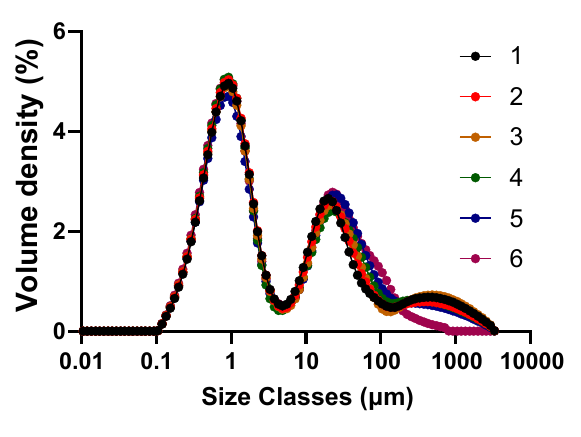}\caption{Size distribution from $n=6$ repeated samples (Malvern Mastersizer).}
\end{subfigure}
\caption{BaSO$_4$ suspension particle characterisation.\label{fig:method1}}
\end{figure}

For the model suspensions, pre-micronised barium sulfate (Sigma-Aldrich) was weighed into the canisters ($2 \pm 0.02$ mg), after which the canisters were crimped and pressure-filled with propellant.
The size and morphology of the BaSO$_4$ particles prior to formulation preparation was analysed using scanning electron microscopy and a Malvern Mastersizer. A sample image and size distribution is shown in Figure \ref{fig:method1}. The suspensions consist of primary particles on the order of 1 \textmu m in agglomerates on the order of 20 \textmu m having an overall $D_{50}$ of $1.73$ \textmu m at 6.7\% relative standard deviation. 

\subsection{Hardware} 

For this study, standard pMDI actuator and canister hardware was used. A Kindeva actuator with 0.3 mm exit orifice diameter was used for the studies with solution formulations, and 0.4 mm exit orifice diameter for the suspension formulations.  Units consisting of 50 \textmu L metering valves (Kindeva Spraymiser with retaining cup) and a FEP-coated aluminium canisters with standard wall thickness were used for all tests \cite{Bradley.2004}.

In order to remotely operate the pMDI units under controlled conditions during X-ray measurements, a specialised testing facility was developed as per Figure \ref{fig:method2}. The actuator was placed into a 3D printed plastic housing which provided a continuous supply of ambient air at 28 std. L/min through the mouthpiece. A small heating element and thermocouple were placed behind the nozzle block in order to stabilise the nozzle block temperature during repeated actuations and prevent icing \cite{DukeRDD2023}. The actuator housing was mounted beneath a pneumatic linear actuator which remotely depressed the unit with a repeatable force and timing. The mouthpiece was fitted into an acrylic spray chamber connected to a continuous flow exhaust duct.  

All units were placed in an ultrasonic bath prior to X-ray measurements, and shaken vigorously by hand for 20 seconds prior to installation. Suspension formulations were shaken between each measurement in order to achieve uniform initial conditions throughout the unit life. Five shots were fired to waste prior to commencement of data collection. A fixed dwell time of 20 s between actuations was used, with data collected after each 10 or 25 shots for solutions and suspensions respectively.

\subsection{X-ray fluorescence experiments}

\begin{figure}
\centering
\begin{subfigure}[b]{.42\textwidth} \centering
	\includegraphics[width=\textwidth]{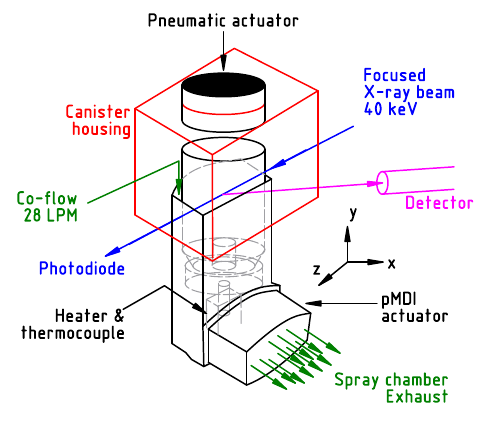}
	\caption{Schematic of experiment.\label{fig:method2}}
\end{subfigure}
\begin{subfigure}[b]{.57\textwidth} \centering
	\includegraphics[width=\textwidth]{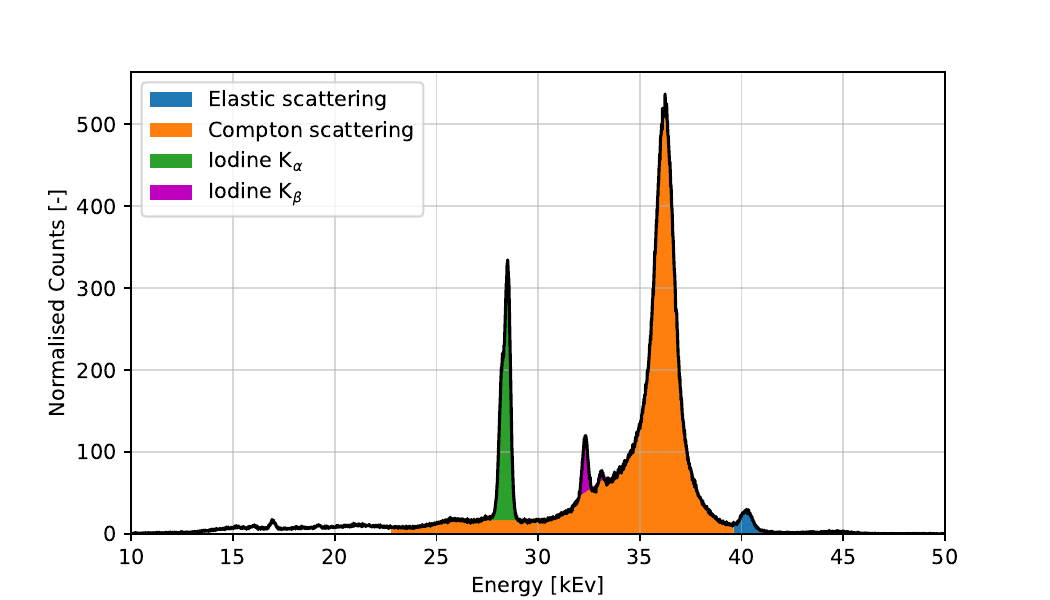}
	\caption{Sample X-ray fluorescence spectrum at the mid-plane of a canister containing 2 mg/mL KI in 8\% w/w ethanol with HFA134a propellant.\label{fig:method3}}\end{subfigure}
\caption{Experiment schematic and sample spectrum.}
\end{figure}

The X-ray fluorescence experiments were conducted at the 7-BM beamline of the Advanced Photon Source (APS) synchrotron radiation facility at Argonne National Laboratory \cite{2012.Kastengrenoi9}. A single bounce of the bending magnet beam from the APS from a water-cooled multilayer mirror was used to produce a monochromatic beam of mean energy 40 keV. This beam was then focused using a pair of Rh-coated Kirkpatrick-Baez mirrors to form focal spot of $6 \times 9$ \textmu m cross-section, which was aligned with the center of the pMDI unit as shown in Figure \ref{fig:method2}. The incident beam intensity $I_0$ was recorded upstream of the test section, and the transmitted beam intensity $I_r$ was recorded using a photodiode placed downstream. The pMDI actuator assembly was translated through the beam in the horizontal ($x$) axis and the vertical ($y$) axis using a pair of servomotor stages.

Excitation of X-ray fluorescence from either iodine (KI solutions) or barium (BaSO$_4$ suspensions) was recorded using a 2 mm thick energy-dispersive silicon drift diode (SDD) detector in back scatter mode on a horizontal plane at approximately 65$^\circ$ relative to the incident beam. The SDD signal was integrated over 2.0 s per measurement to produce a time-average energy spectrum of the emitted radiation at each measurement location in the unit.

A sample spectrum for the KI solution formulation in a full unit at the midpoint of the canister is shown in Figure \ref{fig:method3}.  Due to the large angle of the detector relative to the beam, the number of elastically scattered X-rays (blue) is relatively small. The majority of the signal is comprised of Compton scattering (orange); inelastic scattering due to production of recoil electrons which primarily occurs in the canister wall. The fluorescent emission from Iodine is indicated by the green (K$_\alpha$) and pink (K$_\beta$) emission lines, whose energies are determined by the allowed electron transitions to fill the core shell hole produced when iodine 1s electrons are ejected due to interaction with the incident radiation \cite{2006.Beckhoff}. The K$_\beta$ emission is relatively weaker and contains a strong background due to the underlying Compton scattering, and is thus rejected. The K$_\alpha$ emission peak (green) is integrated and the underlying tail of the Compton scattering signal is removed in order to measure the fluorescent emission intensity $I_{fluor}$.

The signal processing methodolology and underlying calculations for the X-ray fluorescence spectroscopy method have been described in detail in previous work \cite{2019.Duke5w7}. Here, we give a brief overview. The relationship between the local concentration of the API in the unit and the recorded fluorescence signal is determined by the product of:
\begin{itemize} \item the incoming beam flux ($I_0$),
\item the fraction of the beam absorbed by the sample, accounting for the canister, valve, and surrounding environment,
\item the fraction of the absorbed beam that causes ionisation of the fluorescing species, which depends on the average concentration $C$ in the beam path and fluorescence yield $\omega$,
\item the fraction of the fluorescence not re-absorbed by the fluid, canister wall, etc., before reaching the detector,
\item the fraction of the sampling time that the detector is active ($f_d$), and
\item the fraction of the emission solid angle captured by the detector active area $A_{det}$ at distance $r_{det}$.
\end{itemize}
This can be expressed as
\begin{equation}
\frac{I_{fluor}}{I_0} =  \left( 1-\frac{\tau_r}{\tau_{bk}} \right) \, \frac{C \, \omega f_d}{\psi}  \frac{A_{det}}{4 \pi r_{det}^2 \pi}. \label{fluoreq}
\end{equation}

The sample transmission $\tau$ for a monochromatic beam is described by the Lambert-Beer Law
\begin{equation}
\tau = \frac{I}{I_0} = \exp \left( -  \int \mu \rho \, dz \right )\label{eq:rad}
\end{equation}
where $\rho$ is the density and $\mu$ is the attenuation coefficient of the absorbing material (nozzle wall or formulation). A background correction for absorption in the canister wall, valve, seals, 3D printed housing and surrounding air is made by scanning an empty unit prior to the experiment ($\tau_{bk}$). The fraction of the beam absorbed by the fluid inside the canister is determined by the ratio of transmission through the sample in the experiment $\tau_r$ and the empty unit $\tau_{bk}$.

\begin{figure}
\centering
\begin{subfigure}[b]{.49\textwidth} \centering
	\includegraphics[width=\textwidth]{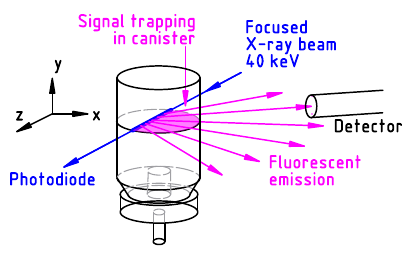}\caption{Canister and X-ray geometry.\label{fig:method4a}}
\end{subfigure}
\begin{subfigure}[b]{.49\textwidth} \centering
	\includegraphics[width=\textwidth,clip=true,trim=3mm 3mm 5mm 1cm]{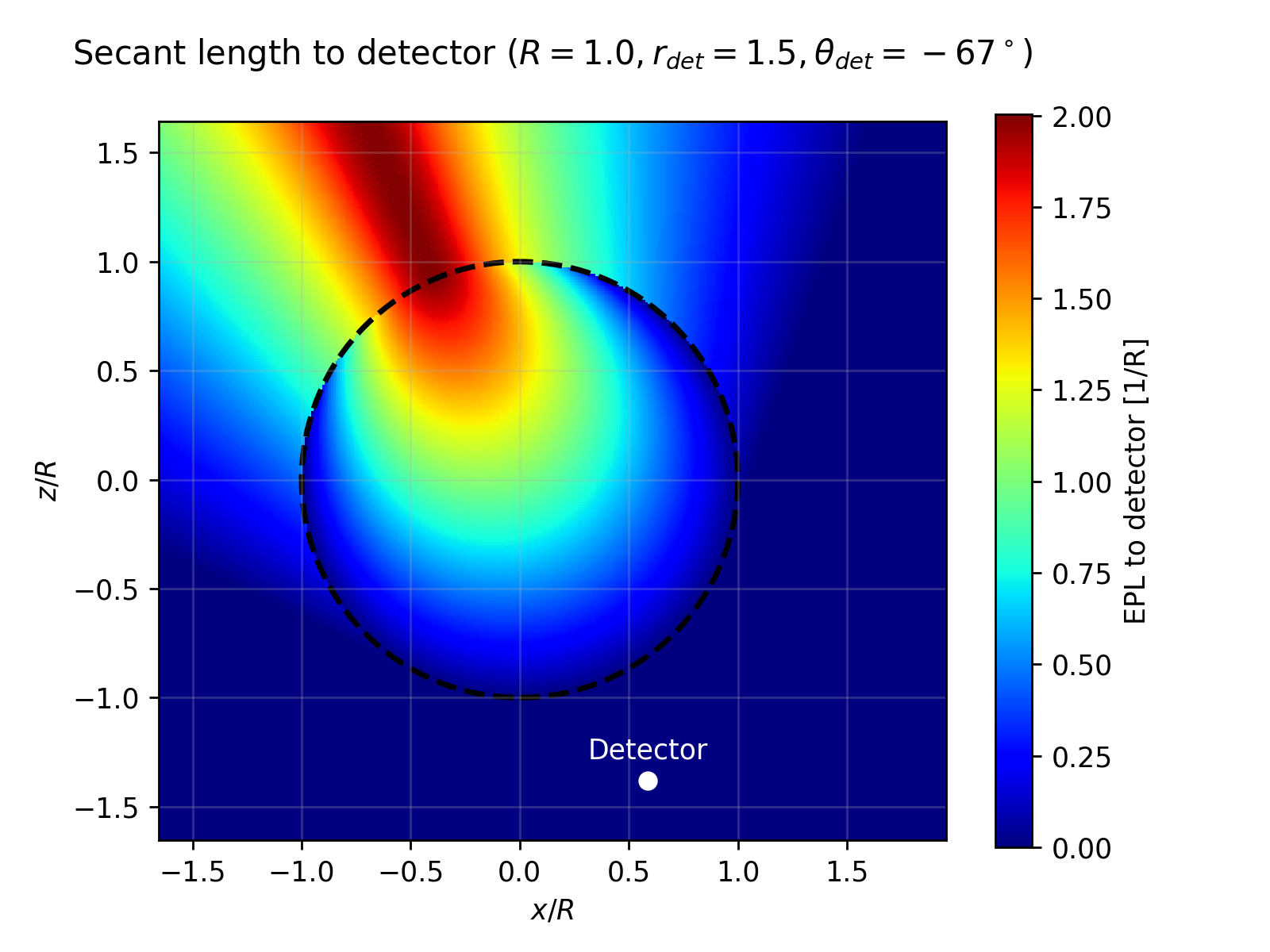}
	\caption{Example signal trapping correction.\label{fig:method4b}}
\end{subfigure}
\caption{X-ray fluorescence geometry and sample signal trapping correction field in a circular cross-section through the canister.\label{method4}}
\end{figure}

The absorption coefficients $\mu$ and fluorescence yields $\omega$ are tabulated for known materials and are obtained from literature \cite{2009.Thompson}. Densities $\rho$ are obtained from NIST data \cite{2007.Lemmon}. $f_d$ is determined from the detector count rate following the method of Walko et al \cite{2011.Walko} and the detector geometry is known based on the experimental setup.

The remaining unknown in Equation \ref{fluoreq} is the signal-trapping coefficient $\psi$. This requires knowledge of the sample geometry. An example is given in Figure \ref{fig:method4a}. The emission of fluorescence (magenta rays) from a line source produced by the incident beam (blue) is partially reabsorbed by the surrounding fluid and canister wall. The magnitude of correction required increases as the amount of material between the beam and the detector increases.  The correction is estimated using a 2D ray-tracing method \cite{2019.Duke5w7} which models the incident beam as a line source in the $z$ axis emitting rays in the $x-z$ plane toward the detector. The recorded intensity is determined by the product of the exponential decay of the incident beam intensity and the transmission of the emitted ray, both of which can be described by equation \ref{eq:rad}. The emitted radiation has a lower energy and $\mu$ is corrected to accomodate this.  The emission calculations considers the product of absorption in the liquid, canister wall, surrounding plastic components, and detector air gap. 

The ray-tracing calculation is demonstrated in Figure \ref{fig:method4b}, for emission at any point in a horizontal plane with a fixed detector position. The canister wall is indicated by the dashed line. The detector is shown much closer to the experiment than its actual position, in order to clearly illustrate the correction method. At the distances used in the experiment (275mm), the change in detector solid angle with measurement position through the canister is sufficiently small to be considered constant. The contribution of rays from points along the beam axis, weighted in $z$ to account for its exponential decay, are summed to calculate an average equivalent path length (EPL) through the sample which can then be related to the absorption once again via the Lambert-Beer Law to obtain $\psi$. Once completed for each measurement position ($x,y$), a projected concentration field in the $z$ axis is obtained. If the local concentration $c$ is given as a mass per unit volume, the measured line-of-sight average along the beam is defined as the integral $C = \int c \, dz$ (units mass per area) \cite{2015.Duke}.

\section{Results and Discussion}

\subsection{Calibration -- solution formulations}

\begin{figure}
\centering
\includegraphics[width=0.75\textwidth]{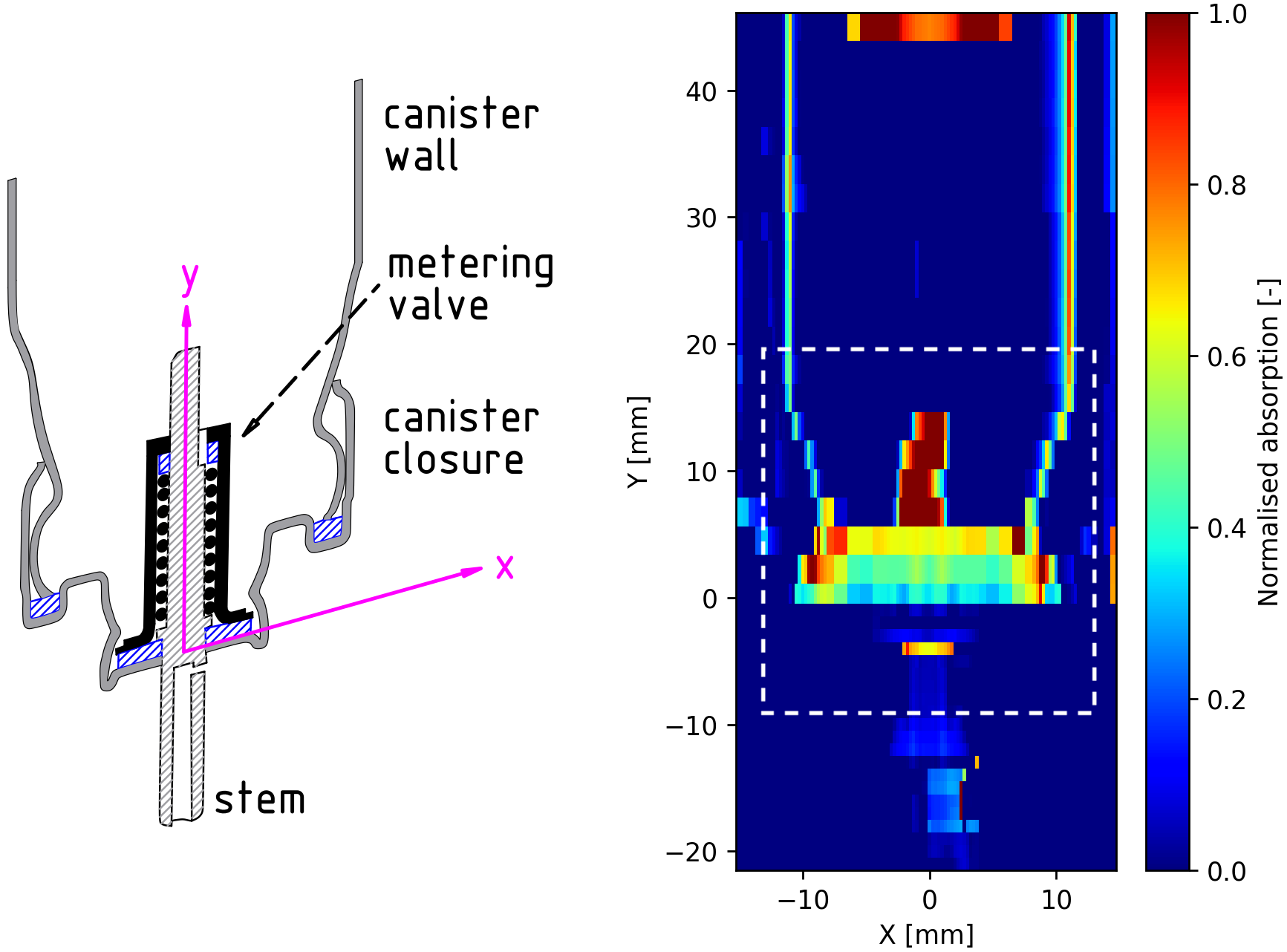}
\caption{Can-valve geometry and corresponding X-ray transmission in an empty unit. Dashed box indicates the region corresponding to the cross-section diagram.\label{fig:results1}}
\end{figure}

Figure \ref{fig:results1} shows a reference absorption field ($1-\tau_{bk}(x,y)$) in an empty unit with the coordinate system origin centered on the valve seat.  It was not possible to obtain a signal from inside the metering valve chamber due to the use of a metal stem. However, the valve closure and gasket region was accessible, as well as the metering valve and canister surfaces.

\begin{figure}
\centering
\begin{subfigure}[b]{0.8\textwidth}
	\includegraphics[width=\textwidth,clip=true,trim=5mm 2mm 2cm 1.4cm]{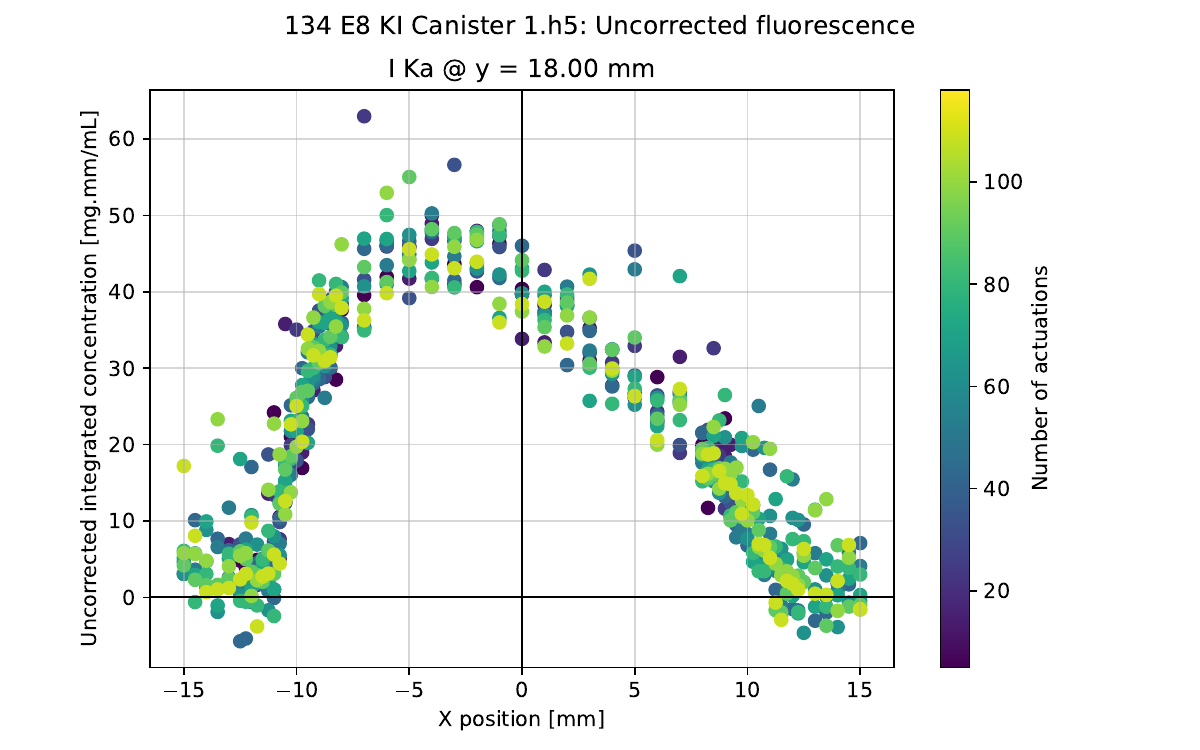}\caption{Uncorrected iodine K$_\alpha$}
\end{subfigure}
\begin{subfigure}[b]{0.8\textwidth}
	\includegraphics[width=\textwidth,clip=true,trim=5mm 2mm 2cm 1.4cm]{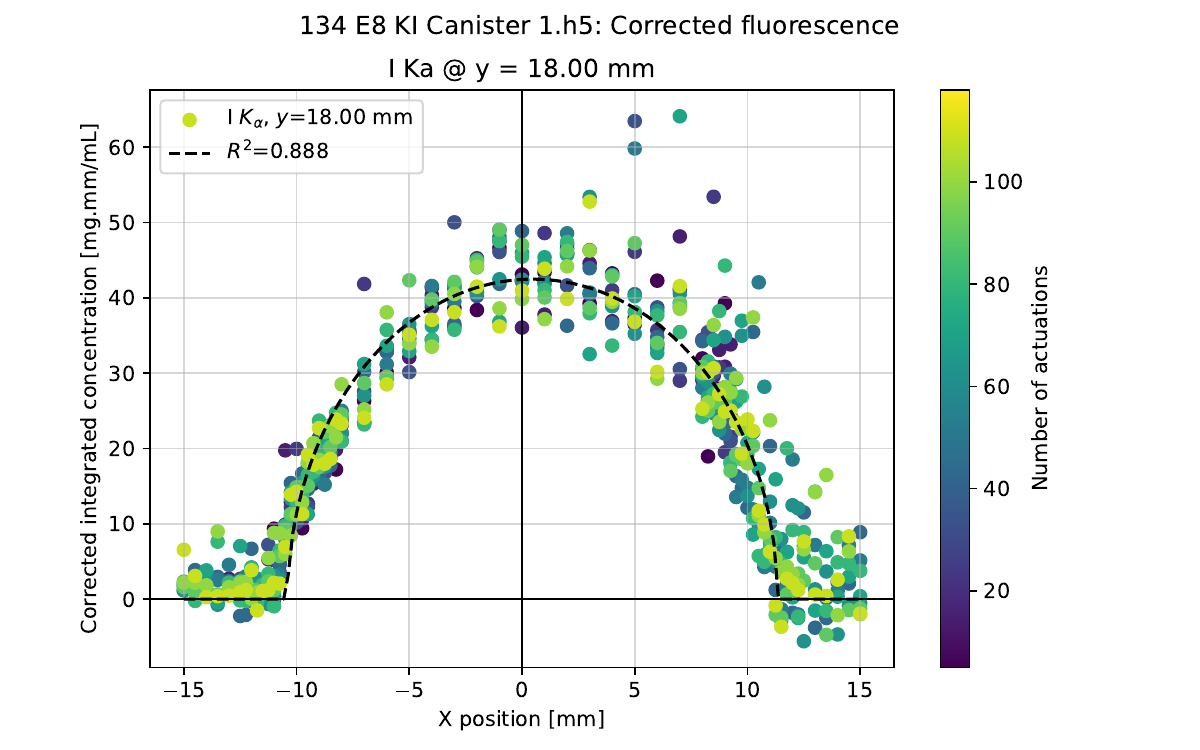}\caption{Corrected iodine K$_\alpha$}
\end{subfigure}
\caption{Calibrated integrated iodine concentration in valve cross-section at $y=18$ mm above the valve (a) without and (b) with signal trapping correction. Sample is 2 mg/mL KI solution in 8\% w/w ethanol and HFA134a propellant. Colour scale indicates number of actuations.\label{fig:results2}}
\end{figure}

A series of calibration studies were completed by scanning HFA134a formulations with a known concentration of 2 mg/mL KI at a fixed height of 18 mm (indicated by uppermost dashed line in Fig. \ref{fig:results1}). This location was selected as it was within the homogeneous liquid region of the bulk formulation away from the valve body. These are shown both before and after the signal-trapping correction (Figure \ref{fig:results2} a \& b respectively). The measurement was repeated between actuations to check for stable solution concentration through the life of the unit. Once the iodine K$_\alpha$ signal was corrected to account for preferential absorption on the far side of the unit ($+x$) an elliptical profile was obtained which was fitted with linear regression to obtain a calibration coefficient to relate the corrected fluorescence intensity to the mass concentration. 

\begin{figure}
\centering
	\includegraphics[width=0.8\textwidth]{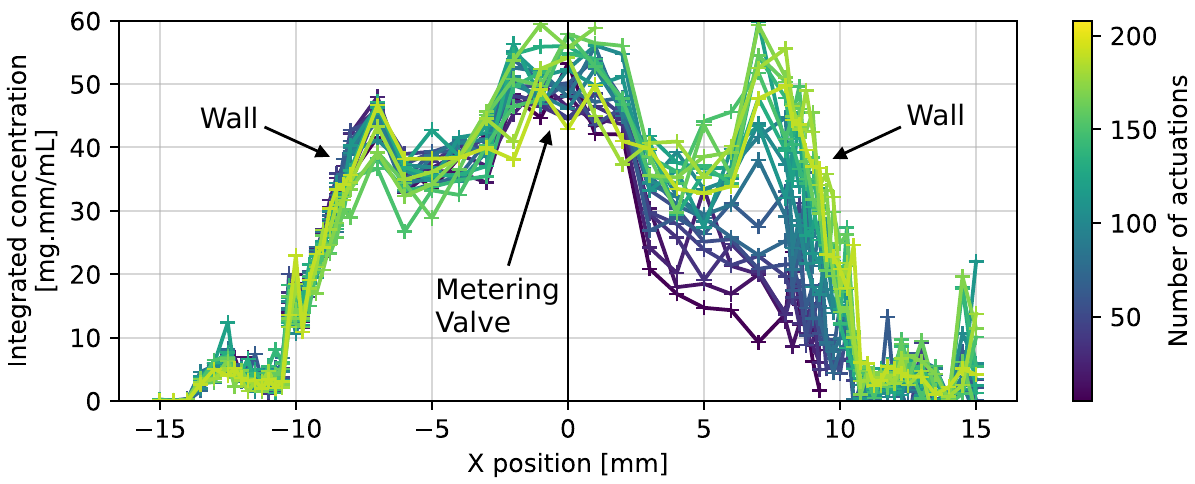}
\caption{Calibrated, correct path-integrated iodine concentration across the canister-valve closure. Sample is 2 mg/mL KI solution in 8\% w/w ethanol and HFA134a propellant. Colour scale indicates number of actuations.\label{fig:results3}}
\end{figure}

The measurement was then repeated in a horizontal plane across the canister-valve closure region; the results are shown in Figure \ref{fig:results3}. A three-peak distribution is observed with higher concentration near the canister wall and metering valve surfaces, and a lower concentration in the bulk liquid. This profile is indicative of an annular distribution with an approximately constant concentration near the canister-valve closure (external wall) and an increased concentration around the base of the metering valve. Through-life performance is indicated by line colour, which represents the number of actuations fired prior to each repeated measurement. As the unit empties, the concentration decreases on the side of the valve facing the metering valve inlet port (right, $x>0$ side, Figure \ref{fig:results3}) while on the opposing (left, $x<0$) side it remains relatively constant.

\subsection{Through unit life performance -- solution formulations}

\begin{figure}[ht]
\centering
\begin{subfigure}[b]{0.49\textwidth} \centering
	\includegraphics[height=5cm]{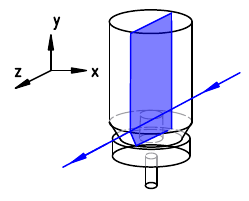}\caption{Measurement plane.\label{fig8a}}
\end{subfigure}
\begin{subfigure}[b]{0.49\textwidth}
	\includegraphics[height=5cm]{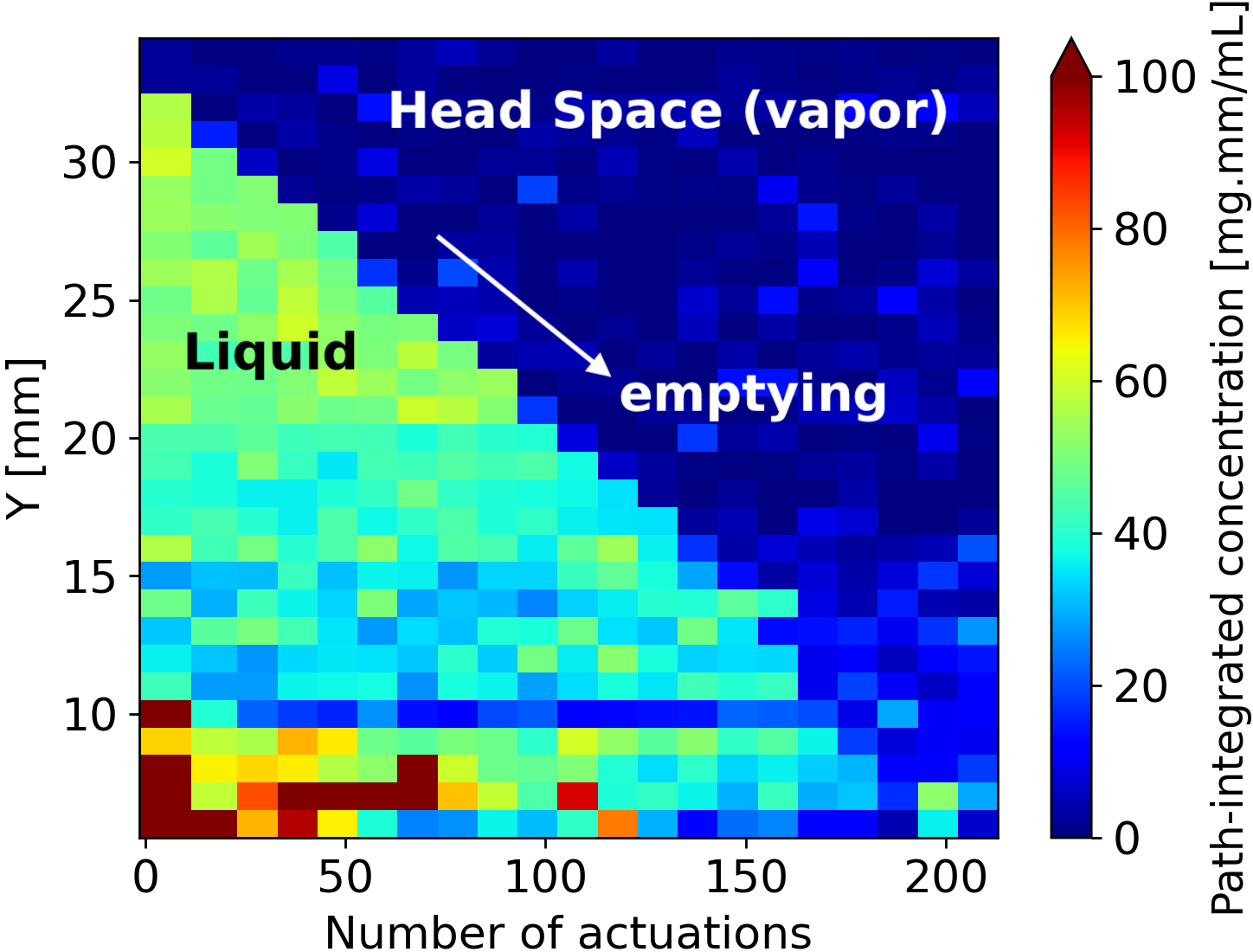}\caption{HFA134a\label{fig8b}}
\end{subfigure}
\begin{subfigure}[b]{0.49\textwidth}
	\includegraphics[height=5cm]{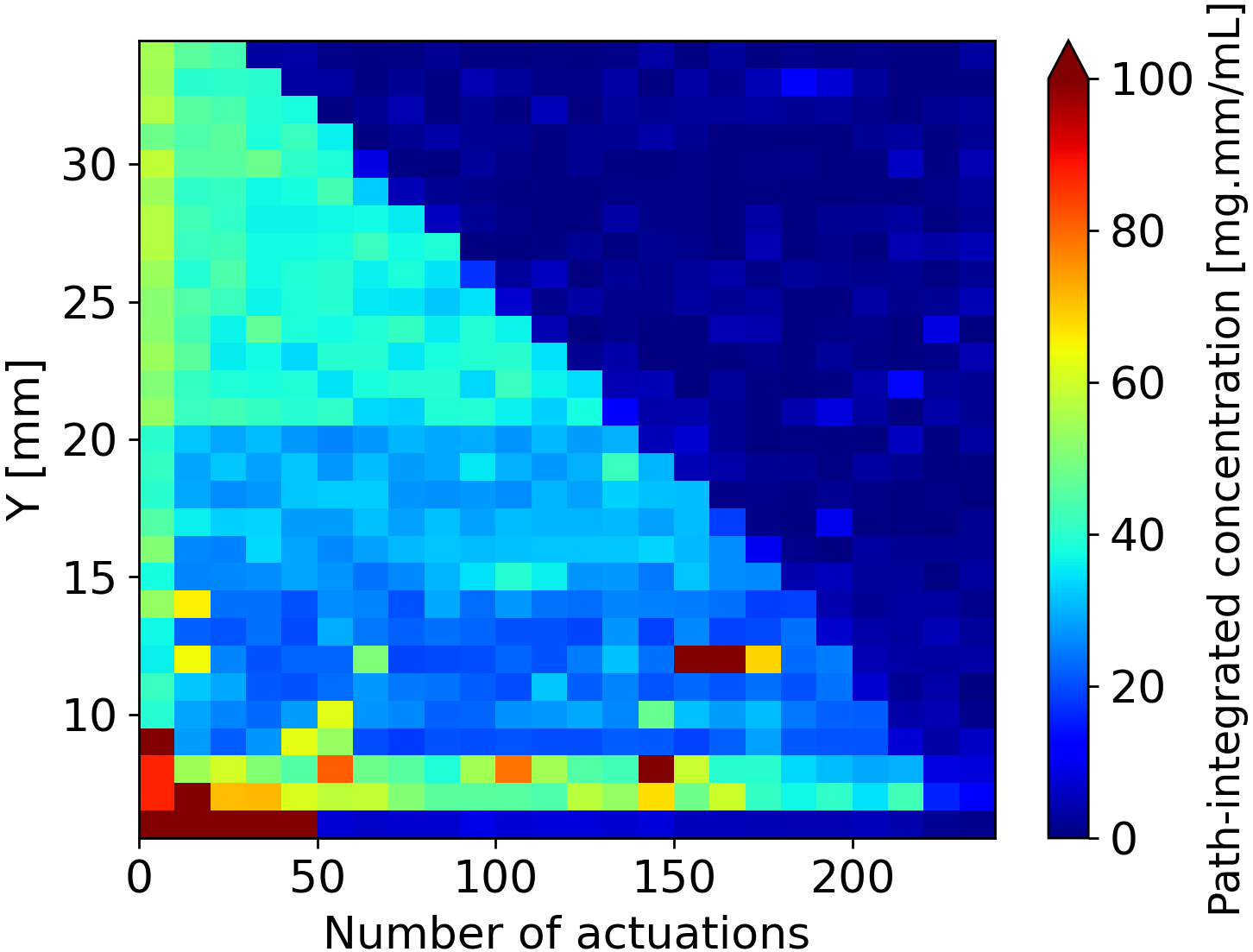}\caption{HFA152a\label{fig8c}}
\end{subfigure}
\begin{subfigure}[b]{0.49\textwidth}
	\includegraphics[height=5cm]{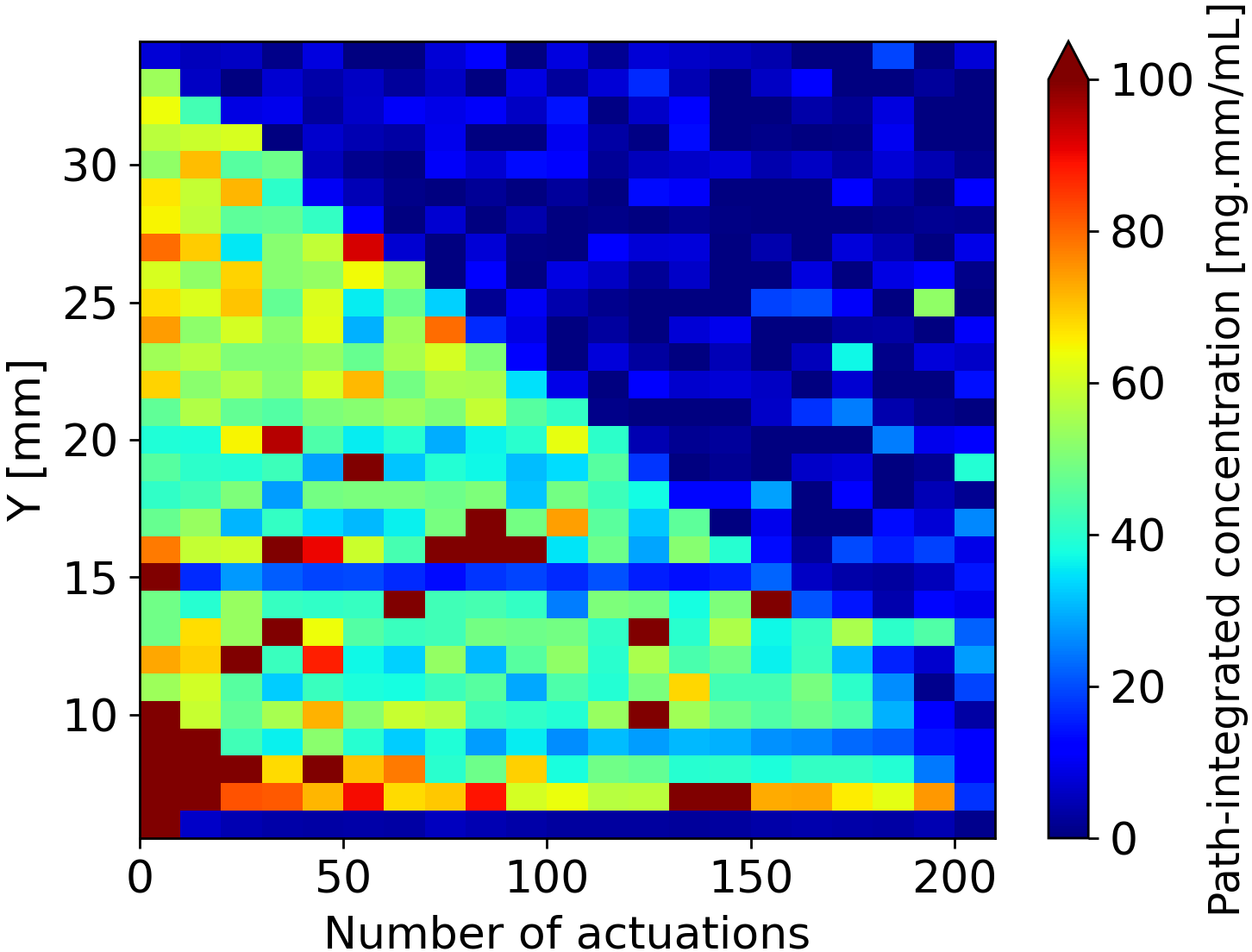}\caption{HFO1234ze(E)\label{fig8d}}
\end{subfigure}
\caption{Path-integrated iodine concentration along the central axis of the canister (a) showing through unit life performance vs. number of actuations (horizontal axis) for (c-d) all propellants with 2 mg/mL KI in 8\% w/w ethanol.\label{fig:results4}}
\end{figure}

In order to assess the through unit life performance of the KI solution formulations, a repeated set of measurements were conducted in the mid plane of the unit as shown in Figure \ref{fig8a}. Units were translated vertically through the beam to develop a vertical ($y$) raster-scan profile. The unit was actuated 10 times, measured with a constant time delay, and this was then repeated until the unit was empty. 

Figures \ref{fig8b}-d show the evolution of this vertical unit KI concentration profile through the unit life. The vertical axis is the vertical distance from the valve with $y=0$ indicating the base of the valve at the seal (see Figure \ref{fig:results1} for reference geometry). The horizontal axis represents the number of actuations fired. The local path-integrated concentration of KI is indicated by the colour scale. The expected equilibrium concentration for a homogenised mixture at the mid plane of the unit corresponds to green, with blue indicating low concentration and red indicating above-average concentration. The liquid fill line is visible as the transition from green to blue in Figures \ref{fig8b}-d. 

Two features of interest are observed in the vertical profiles in Figure \ref{fig:results4}. The first are horizontal striations which are particularly evident for HFA134a and HFA152a (Figs. \ref{fig8b} \& \ref{fig8c}) below $y=20$ mm. These remain once the background effects of the canister wall and surrounding plastic and metal components are normalized out.  The position of these striations varies from one canister to the next. 
They are likely caused by precipitate KI crystals that sink in the formulation and accumulate near the bottom of the unit, but this was not verified.

\begin{figure}
\centering
\includegraphics[width=\textwidth,clip=true,trim=2.5cm 4.5cm 7mm 2.5cm]{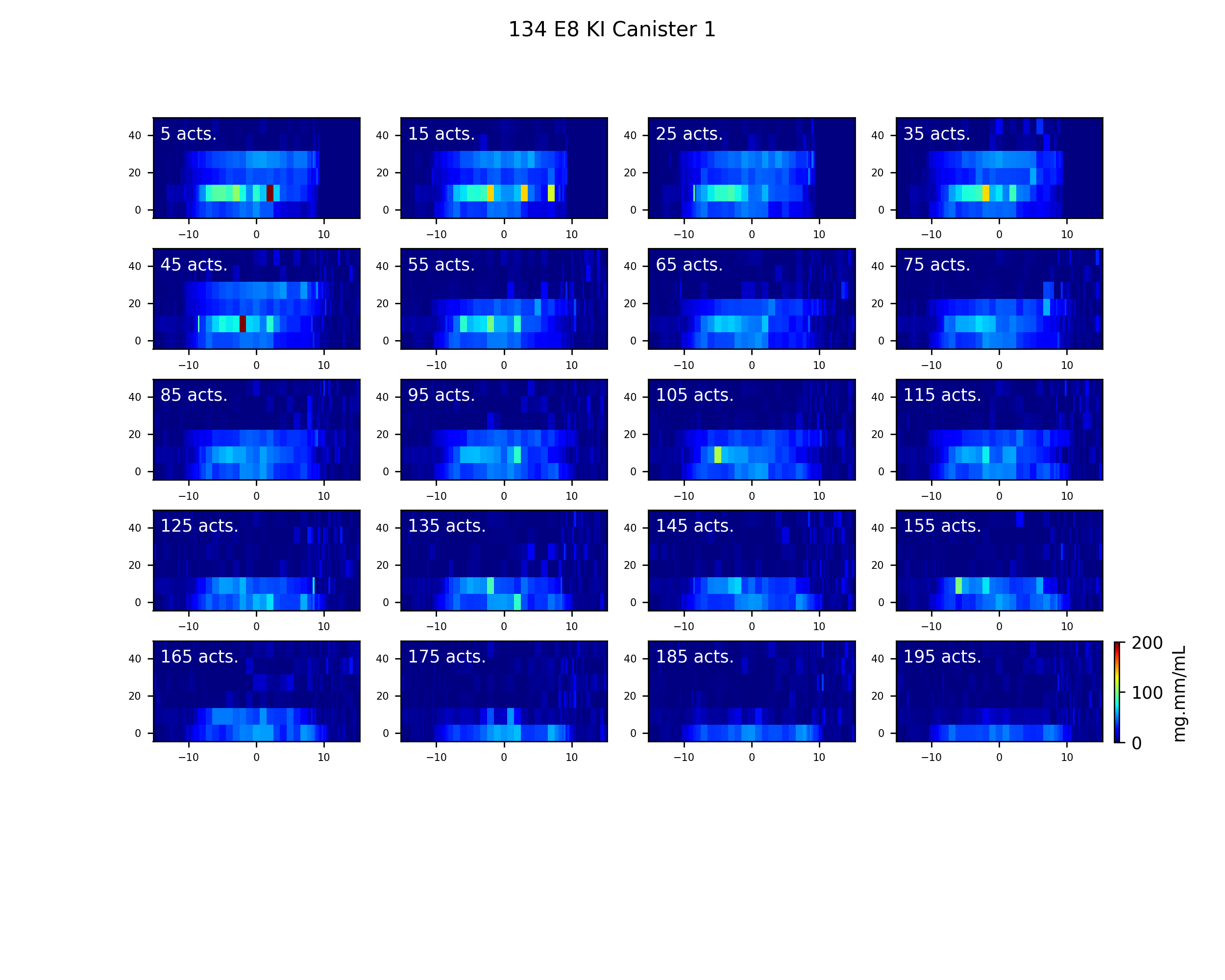}
\caption{Through unit life canister-valve closure deposition for 2 mg/mL KI in 8\% w/w ethanol in HFA134a propellant. Coordinates are given in mm relative as per Fig. \ref{fig:results1}.
\label{fig:results6}}
\end{figure}

\begin{figure}
\centering
\includegraphics[width=\textwidth,clip=true,trim=2.5cm 4.5cm 7mm 2.5cm]{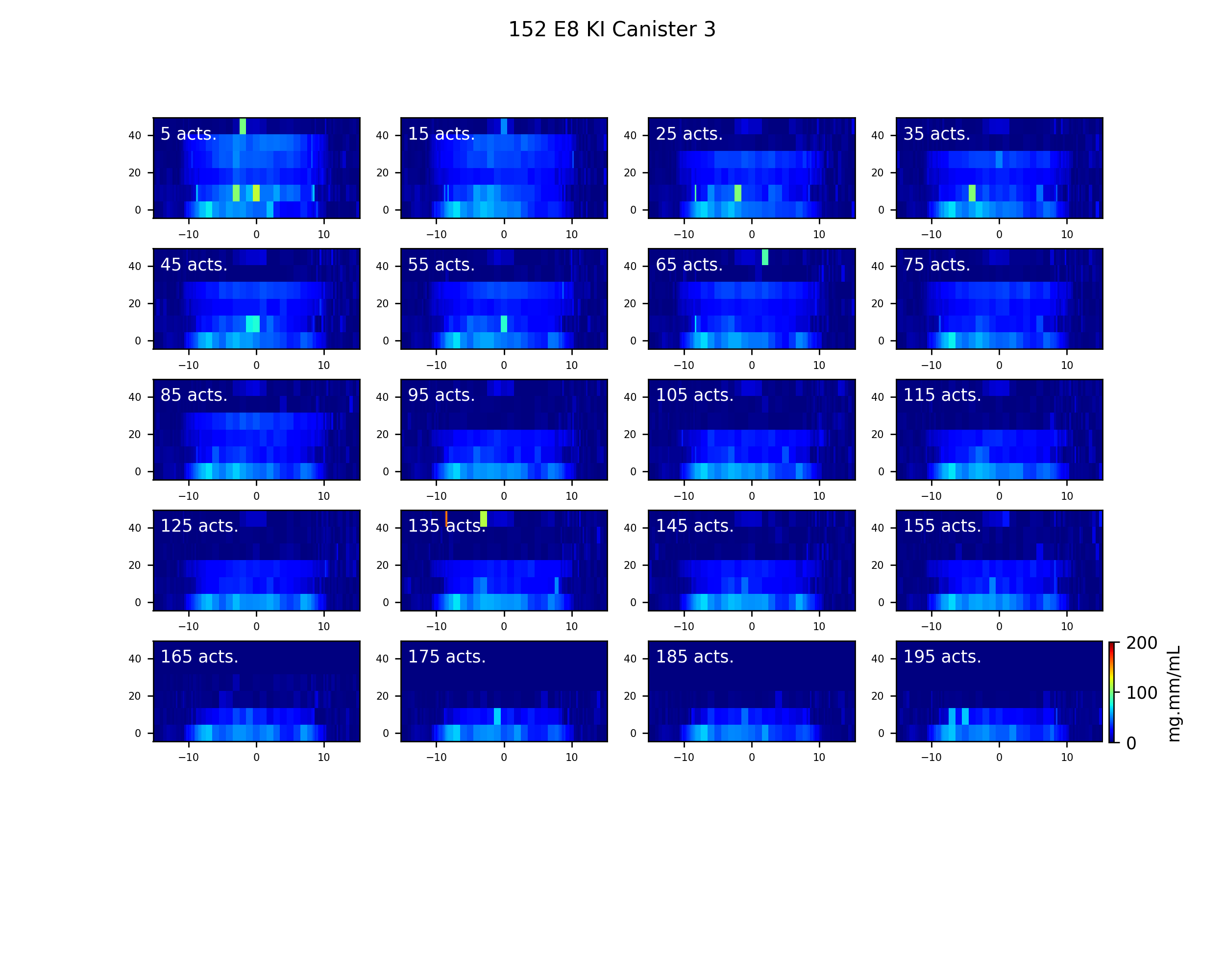}
\caption{Through unit life canister-valve closure deposition for 2 mg/mL KI in 8\% w/w ethanol in HFA152a propellant. Coordinates are given in mm relative as per Fig. \ref{fig:results1}.
\label{fig:results7}}
\end{figure}

\begin{figure}
\centering
\includegraphics[width=\textwidth,clip=true,trim=2.5cm 4.5cm 7mm 2.5cm]{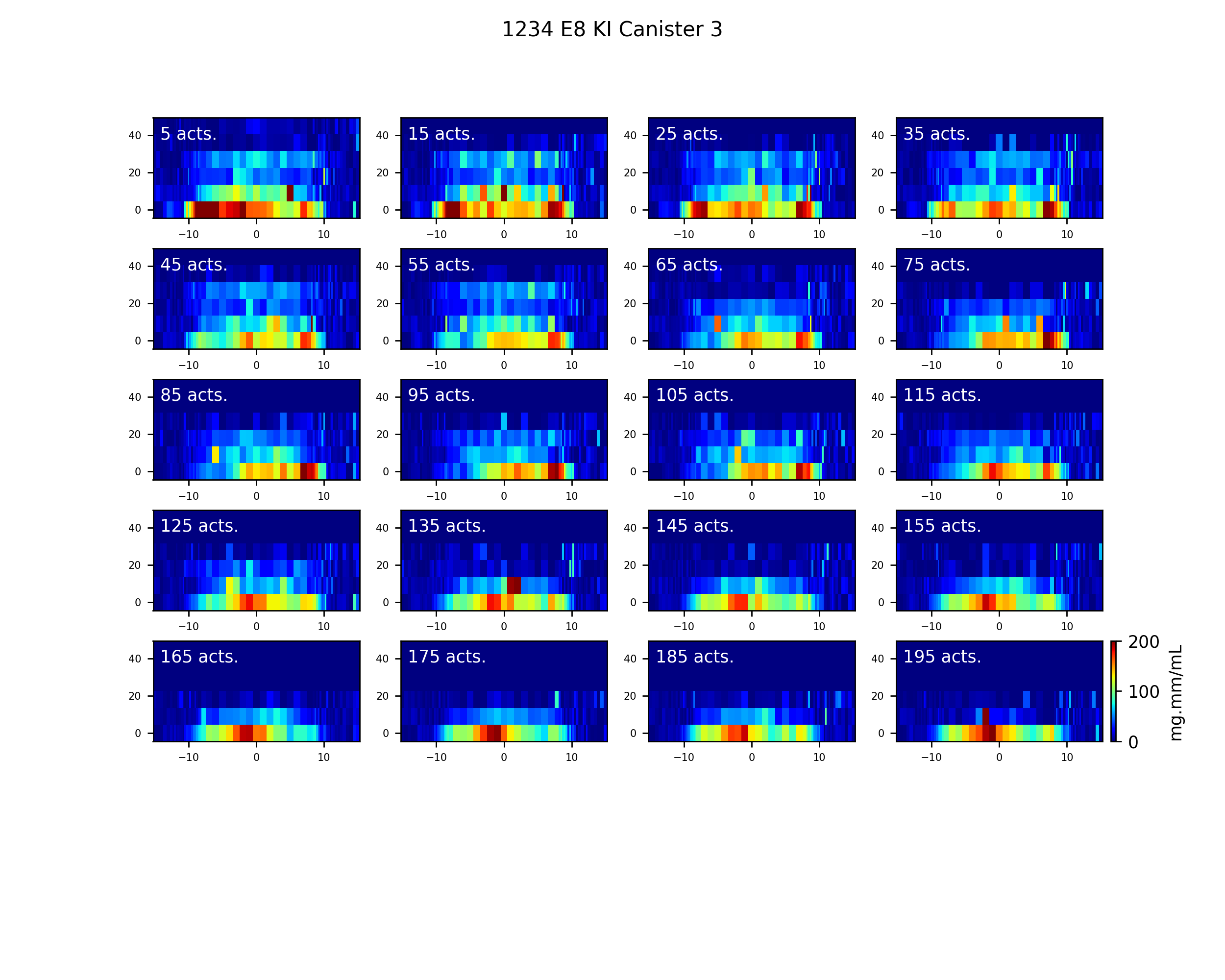}
\caption{Through unit life canister-valve closure deposition for 2 mg/mL KI in 8\% w/w ethanol in HFO1234ze(E) propellant. Coordinates are given in mm relative as per Fig. \ref{fig:results1}.
\label{fig:results8}}
\end{figure}

The second feature of interest in Figure \ref{fig:results4} is the presence of high concentrations below $y=10$ mm (i.e. in the valve closure and valve seal region).  These are present for all formulations, but are more pronounced for HFO1234ze(E).  This was investigated in further detail by scanning the canister-valve closure region in both $x$ and $y$ axes to develop a 2D fluorescence map of KI deposition every 10 actuations through unit life. These profiles are shown for HFA134a, HFA152a and HFO1234ze(E) propellants respectively in Figures \ref{fig:results6}, \ref{fig:results7} \& \ref{fig:results8} with a uniform colour scale. Deposition is maximum at start of unit life and decreases slightly with use. The deposition in HFO1234ze(E) formulations (Figure \ref{fig:results8}) is substantially higher than for HFA formulations (Figs. \ref{fig:results6}-\ref{fig:results7}).

\begin{figure}
\centering
	\includegraphics[height=8cm,clip=true,trim=9cm 0 0 0]{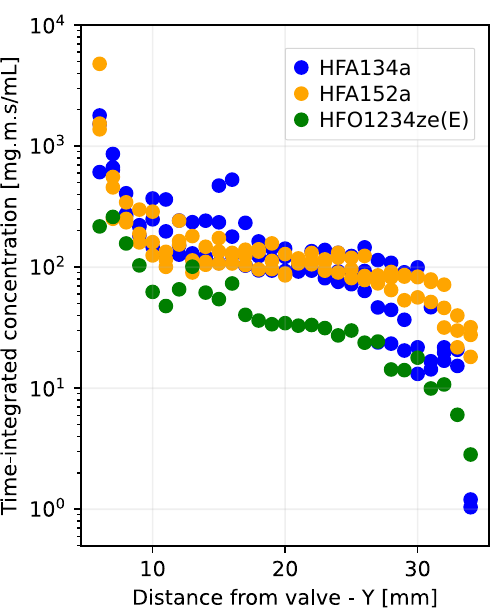}
\caption{Integrated iodine concentration along the central axis of the unit for all propellants with 2 mg/mL KI in 8\% w/w ethanol.\label{fig:results5}}
\end{figure}

In order to quantitatively assess differences in the distribution of the API between formulations, the spatial profiles were integrated to assess total API content. Integration of the path-integrated concentration in the vertical axis over time yields an integral $\int C(y) dt = \iint c(y,z) \, dz \, dt$
which is shown for all formulations in Figure \ref{fig:results5} as a function of vertical distance from the valve. For a homogeneous mixture and constant metering valve volume, we expect this profile to decrease linearly with number of actuations due to unit emptying. 
However, all formulations show larger values near the valve ($y < 10$ mm) and a plateau region at the midpoint ($y \approx 20$ mm). This is due to both sinking of precipitates (which cause very high values at $y \approx 0$, and preferential deposition around the canister-valve closure which causes high values for $1 < y < 10$ mm. This deposition may be attributed to both the lack of coating on the valve components, and increased surface area in this region.
The effect is consistent for HFA134a and HFA152a formulations, but the HFO1234ze(E) formulation is again an outlier with notably lower integrated values in the bulk liquid, and  higher values near the valve. 
This combination of high canister-valve closure measurements (Fig. \ref{fig:results8}) and low volume-integrated measurements (Fig. \ref{fig:results5}) is counterintuitive. However, since the total mass of KI is conserved in the sample, a reduction in the spatially integrated quantity with larger local peak values indicates that for HFO1234ze(E) a significantly larger fraction of KI has precipitated out of solution or deposited onto the valve surface than expected, and is no longer in the bulk fluid phase.

\subsection{Suspension formulations}

\begin{figure}
\centering
\begin{subfigure}[b]{0.49\textwidth} \centering
	\includegraphics[width=\textwidth,clip=true,trim=3mm 2mm 1.3cm 1.4cm]{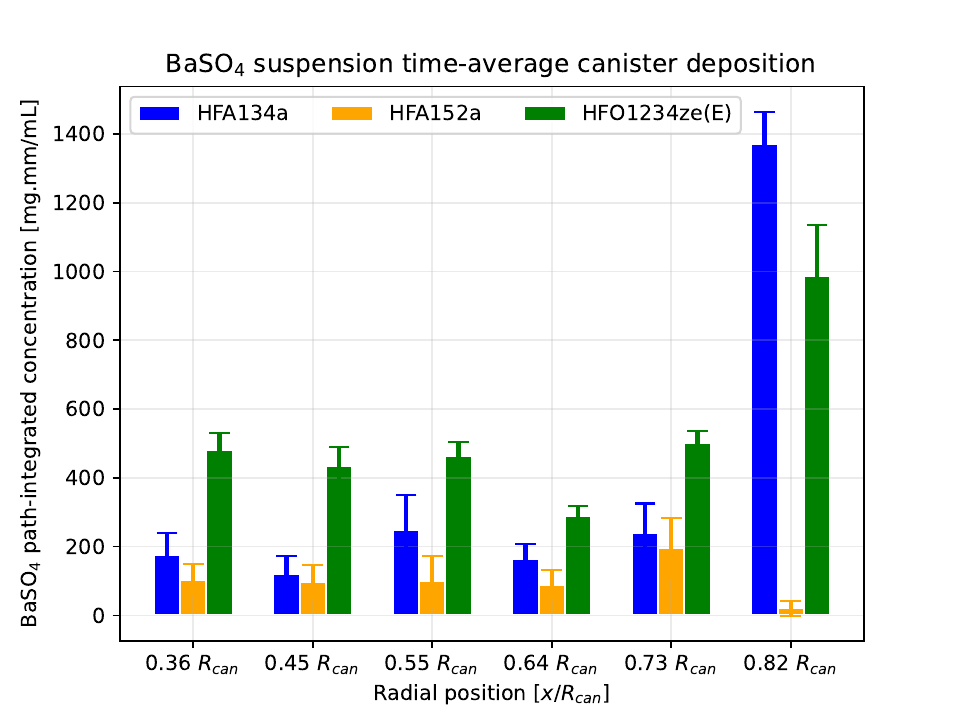}\caption{Radial profiles ($x$).\label{fig:results9a}}
\end{subfigure}
\begin{subfigure}[b]{0.49\textwidth} \centering
	\includegraphics[width=\textwidth,clip=true,trim=3mm 2mm 1.3cm 1.4cm]{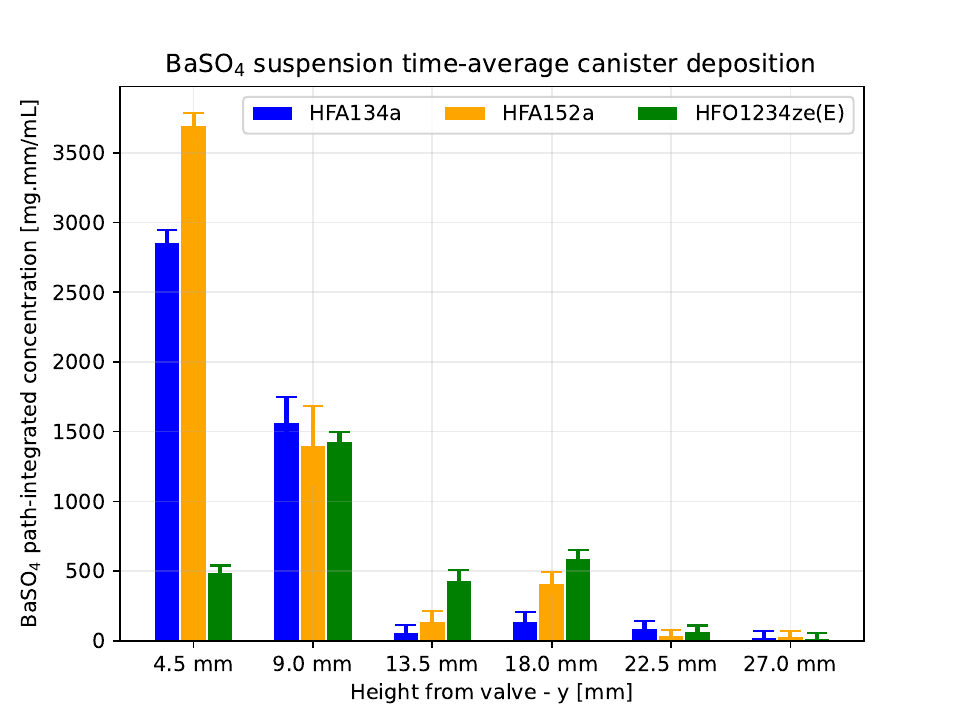}\caption{Axial profiles ($y$).\label{fig:results9b}}
\end{subfigure}
\caption{Integrated barium concentration along (a) horizontal and (b) vertical planes, showing through life performance vs. number of actuations (horizontal axis) with 2 mg/mL suspended BaSO$_4$ in pure propellant.\label{fig:results9}}
\end{figure}

Due to the short sedimentation timescales for the BaSO$_4$ suspension formulations owing to their high density, data was collected by shaking the unit, fixing the unit at a single measurement position, and sampling the fluorescence signal repeatedly at a single position to obtain a time-series. The measurements were repeated over a grid of positions through the unit canister and valve. Figure \ref{fig:results9} shows the time-integrated BaSO$_4$ concentration for all three propellants.

Figure \ref{fig:results9a} shows the variation with radial position, which is normalized on the horizontal axis with respect to the unit radius. Both HFA134a and HFO1234ze(E) formulations show high concentrations at larger radii indicative of deposition on the walls occurring at similar time scales. HFA152a suspension formulations have relatively low values near the canister wall indicating that the low density of the propellant is driving particles to sediment rather than adhere to the inner surfaces. 

Figure \ref{fig:results9b} shows the variation with vertical position (distance from the valve). Here, we see the effect of sedimentation for both HFA formulations with high concentrations at $y=4.5$ mm (canister-valve closure region).  In the vertical profile, HFO1234ze(E) is the outlier with higher BaSO$_4$ concentration around the valve and less in the canister closure region, suggesting preferential deposition onto the walls and canister-valve closure above the seal.

\subsection{Actuator deposition}
\begin{figure}
\centering
\begin{subfigure}[b]{\textwidth} \centering
	\includegraphics[width=\textwidth,clip=true,trim=2cm 1cm 1cm 1.8cm]{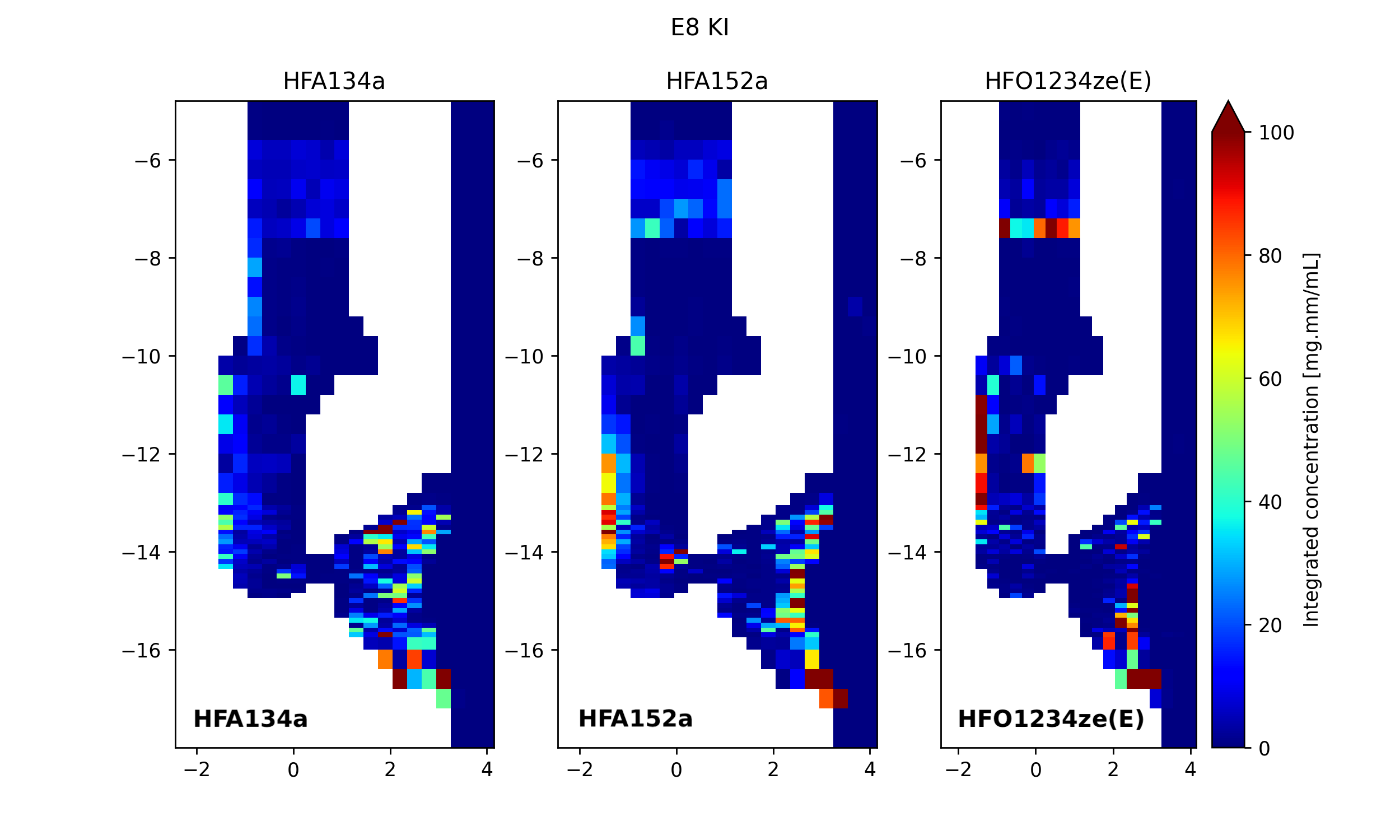}\caption{2 mg/mL KI solution in 8\% w/w ethanol.\label{fig:results10a}}
\end{subfigure}
\begin{subfigure}[b]{\textwidth} \centering
	\includegraphics[width=\textwidth,clip=true,trim=2cm 1cm 1cm 1.8cm]{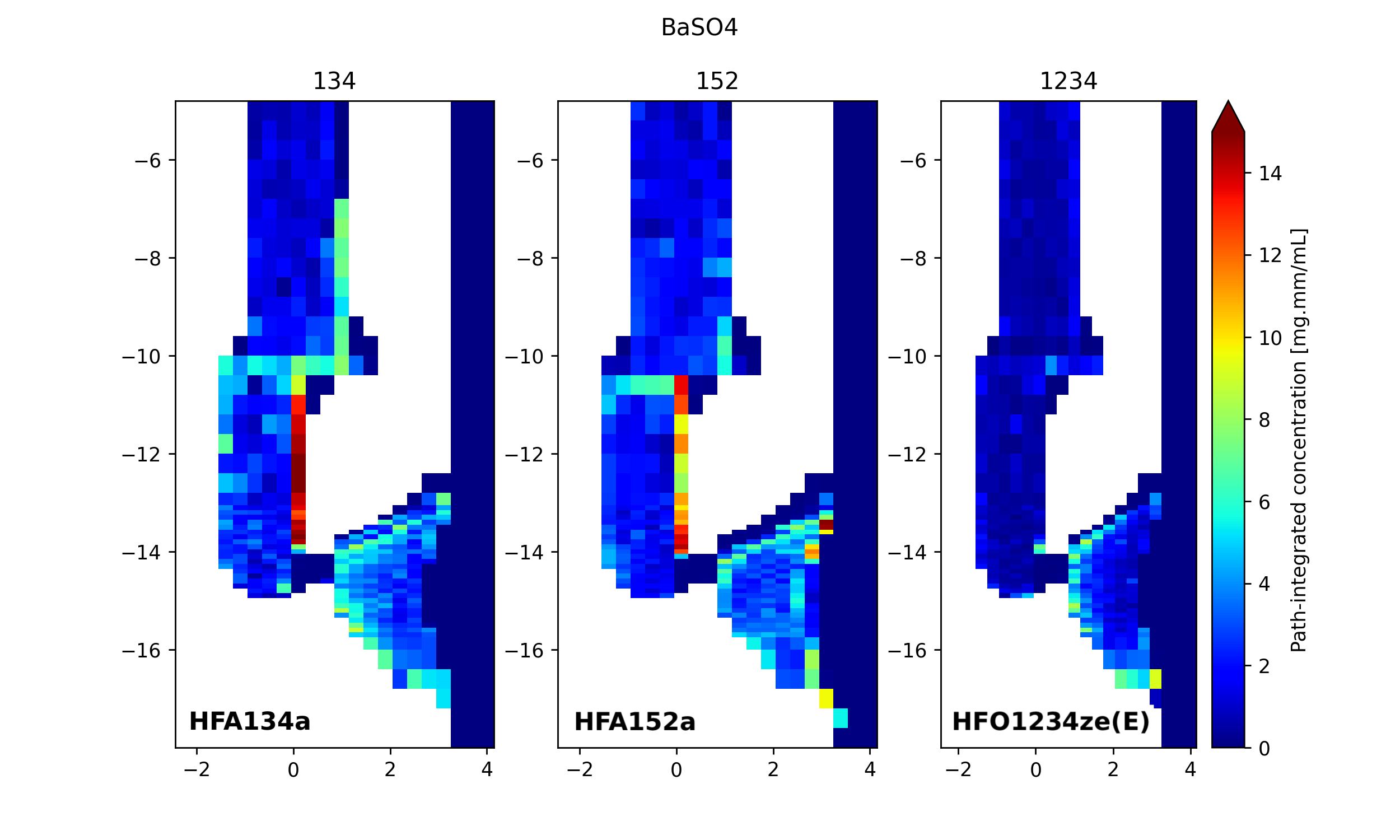}\caption{2 mg/mL BaSO$_4$ suspension.\label{fig:results10b}}
\end{subfigure}
\caption{Valve stem and actuator deposition after 50 actuations for (a) solution and (b) suspension formulations. Axes coordinates in mm.\label{fig:results10}}
\end{figure}

For all formulations, raster-scan measurements were repeated in the stem, actuator expansion chamber and orifice after 50 actuations to assess relative deposition in these components. Results for KI solution formulations with 8\% w/w ethanol are shown in Figure \ref{fig:results10a}, and results for BaSO$_4$ suspensions in \ref{fig:results10b}. Significantly less actuator deposition was observed for the suspension formulations (note colour scale in Fig. \ref{fig:results10a} is 6.7 times larger than Fig. \ref{fig:results10b}). Increased deposition in solution formulations is likely due to residual ethanol in the actuator sump and orifice after the prior actuation, which leaves an API-rich deposit that then influences the subsequent spray \cite{2015.Duke}.  A small reduction in deposition was also observed in HFO1234ze(E) formulations relative to HFA formulations, likely due to reduced concentration of API in this formulation, owing to the increase in deposition on the canister wall and valve.

\section{Conclusions}

In this paper we have demonstrated the efficacy of X-ray fluorescence spectroscopy as a non-invasive and non-destructive \textit{in situ} measurement technique for assessing canister wall, valve and closure deposition in metered-dose inhaler formulations.  A comparison of HFA134a, HFA152a and HFO1234ze(E) propellants with a model solution formulation indicate a notable increase in HFO1234ze(E) canister valve closure deposition, with a corresponding decrease in volume-integrated concentration in the bulk fluid above the valve. Variations in solution formulation concentration are revealed by the sensitivity of X-ray fluorescence to dissolved tracer species, which may not be visible in transparent vials.

Measurements in in pMDI units of sedimenting suspension formulations show that the denser HFA134a and HFO1234ze(E) formulations have increased unit wall deposition relative to rapidly-sedimenting HFA152a formulations. Both HFA formulations indicate sediment buildup around the valve seal and canister valve closure, as expected, but HFO1234ze(E) formulations show far less sediment in this region, with the balance being deposited at the canister wall and valve surfaces.

Results obtained with HFO1234ze(E) propellant suggest that changes in intermolecular attractive forces may play a significant role in reducing the solubility of small molecules in HFO relative to HFA propellants. 
Increased deposition on canister walls and internal unit surfaces occur in both solution and suspension HFO formulations. These changes correlate with the lower dipole moment of HFO1234ze(E) (Table \ref{propellantPropertyTable}) and its altered polarizability \cite{Sampson.2019}. As with HFA and HFC MDI formulations, the mechanism driving these changes is likely to be complex and difficult to predict \cite{1999.Vervaet}. The accumulation of precipitates on the internal surfaces may explain the concentration variations observed in the KI solutions, which are more pronounced in HFO formulations. Conversely, competition between buoyant forces (i.e. sedimentation) and wall interactions are a more likely driver of variations in suspension formulations. In making these hypotheses we note that API interactions with HFO formulations are not presently well understood. At present there is no direct evidence of changes in partitioning into elastomeric valve gasket materials, but this requires further investigation. 

The changes in deposition observed in pMDIs formulated with HFO1234ze(E) propellant are statistically significant but not so large as to be prohibitive. As with the CFC to HFA transition, adjustments to drug dosing levels may be required to accomodate increased internal unit losses due to deposition \cite{1999.Vervaet}. Canister coatings and valve gasket material selection will also need to be considered carefully \cite{Suman.2014,Ashayer.2004}. End user compliance with correct shaking procedures will become even more important in low-GWP formulations \cite{Hatley.2017}. The effect of pMDI product shelf life will be a matter of particular concern \cite{Lewis.2023}. Non-invasive in situ measurement techniques such as X-ray fluorescence spectroscopy provide a useful alternative to optical analysis methods when investigating such matters.

\section*{Acknowledgements}

The authors acknowledge the support of the Australian Research Council (grants LP190100938 \& DP200102016). Use of the Advanced Photon Source, an Office of Science User Facility operated for the U.S. Department of Energy (DOE) Office of Science by Argonne National Laboratory, was supported by the U.S. DOE under Contract No. DE-AC02-06CH11357. The authors thank Brandon Sforzo and Aniket Tekawande (Argonne National Laboratory) for their support. We gratefully acknowledge the computing resources provided on \textit{Bebop}, a high-performance computing cluster operated by the Laboratory Computing Resource Center at Argonne National Laboratory. The authors also acknowledge the use of the National Computational Infrastructure (NCI), which is supported by the Australian Government.

\section*{CRediT author statement}
\textbf{Daniel Duke:} Conceptualization, Methodology, Investigation, Software, Writing-Original draft preparation, Funding acquisition.
\textbf{Lingzhe Rao:} Investigation, Formal analysis.
\textbf{Alan Kastengren:} Methodology, Investigation, Validation, Writing - Review \& Editing.
\textbf{Benjamin Myatt:} Resources, Writing - Review \& Editing, Project administration.
\textbf{Phil Cocks:} Resources, Writing - Review \& Editing, Project administration.
\textbf{Stephen Stein:} Resources, Writing - Review \& Editing, Project administration.
\textbf{Nirmal Marasini:} Investigation, Formal analysis.
\textbf{Hui Xin Ong:} Resources, Supervision, Writing - Review \& Editing.
\textbf{Paul Young:} Resources, Supervision, Writing - Review \& Editing.



\bibliography{xrf23_preprint}

\end{document}